\documentclass[]{spie}  %>>> use for US letter paper
\usepackage{amsmath,amsfonts,amssymb}
\usepackage{graphicx}
\usepackage[colorlinks=true, allcolors=blue]{hyperref}
\usepackage{fancyvrb}

\title{Simulation tools for realistic high-order wavefront correction with the Roman Coronagraph}

\author[a]{A. Lau}
\author[b]{A. J. Riggs}
\author[b]{S. F. Redmond}
\author[b]{E. Cady}
\author[b]{J. Krist}
\author[c]{I. Laginja}
\author[d]{D. Sirbu}
\author[e]{L. Delaye}
\author[f]{R. Soummer}
\author[f]{S. Noiret}
\author[a,g]{J. C. Lambert}
\author[f]{A. Bidot}
\affil[a]{Aix Marseille Université, CNRS, CNES, LAM, Marseille, France}
\affil[b]{Jet Propulsion Laboratory, California Institute of Technology, USA}
\affil[c]{Université Côte d'Azur, Observatoire de la Côte d'Azur, CNRS, Laboratoire Lagrange, Nice, France}
\affil[d]{NASA Ames Research Center, Moffett Field, CA 94035, USA}
\affil[e]{LIRA, Observatoire de Paris, Université PSL, Sorbonne Université, Université Paris Cité, CY Cergy Paris Université, CNRS, 92195 Meudon, France}
\affil[f]{Space Telescope Science Institute, 3700 San Martin Drive, Baltimore, MD 21218, USA}
\affil[g]{Centre de donnees Astrophysiques de Marseille (CeSAM), France}

\authorinfo{Further author information, send correspondence to A. L.: alexis.lau@lam.fr}

\usepackage{xcolor}

\newcommand{\FIG}[1]{{Fig.~}}
\newcommand{\EQ}[1]{{Eq.~}}
\newcommand{\SECT}[1]{{Sect.~}}

\begin{document} 
\maketitle

\begin{abstract}
The Roman Space Telescope Coronagraph Instrument (Roman CGI) will demonstrate high-contrast imaging from space using coronagraphic masks, deformable mirrors, and high-order wavefront sensing and control (HOWFSC). After the baseline technology demonstration, the Roman Coronagraph Community Participation Program (CPP) will pursue science and engineering studies that require realistic predictions of instrument behaviour, observing efficiency, and wavefront control performance. This paper describes \texttt{corgihowfsc}, a configurable simulation framework for repeatable Roman CGI HOWFSC studies. The framework preserves the Roman ground-in-the-loop (GITL) workflow represented by the NASA \texttt{cgi-howfsc} package, while allowing the images to be generated by the higher-fidelity \texttt{corgisim} model. In this configuration, \texttt{cgi-howfsc} remains the reference implementation for estimation, control, and compact-model Jacobian generation; \texttt{corgisim} can supply more flight-like images for studies of instrument performance and robustness. \texttt{corgihowfsc} also coordinates exposure planning, camera settings, expected iteration timing, and contrast normalisation of the HOWFSC loop through \texttt{cgi-eetc}, calibration-related workflows through \texttt{cgi-coralign}, structured diagnostics, and local or distributed execution. By exposing observing modes, image models, probe choices, estimators, controllers, deformable-mirror settings, and runtime options through reusable configuration files, \texttt{corgihowfsc} enables controlled comparisons between reference compact-model simulations and higher-fidelity HOWFSC studies.
\end{abstract}

% Include a list of keywords after the abstract; whatever we are using on the website is fine

\section{INTRODUCTION}
\label{sec:intro}  
The Roman Coronagraph Instrument (Roman CGI) \cite{Mennesson2022, Bailey2023} is designed as a technology demonstration for future space-based direct imaging of exoplanets and circumstellar discs. Its top-level requirement is to demonstrate a broadband dark hole (DH): a selected region of the image in which starlight is suppressed to enable faint companion or disc detections. The instrument creates this DZ using high-order wavefront sensing and control (HOWFSC) with deformable mirrors (DMs) \cite{Cady2025}. In a typical HOWFSC loop, small predefined DM patterns
known as probes are applied to estimate the complex electric field in the DZ \cite{Giveon2011SPIE}; DM commands are computed from that estimate, and the process is iterated until the desired contrast is reached or the control sequence is complete.

For Roman CGI, this control loop is planned as a ground-in-the-loop (GITL) workflow \cite{Cady2025}: images acquired by the instrument are sent to the ground, the HOWFSC software estimates the electric field and computes the DM updates, and the resulting commands are uplinked for the next control step. Therefore, performance studies should preserve the same ground-control that will be used for instrument operations. The NASA \texttt{cgi-howfsc} \cite{cgihowfsc}package\footnote{\url{https://github.com/nasa-jpl/cgi-howfsc}} , a dependency of \texttt{corgihowfsc}\footnote{\url{https://github.com/roman-corgi/corgihowfsc}}, represents this reference HOWFSC workflow, including the compact optical model used for control and Jacobian generation.

The Community Participation Program (CPP) \cite{Savransky2024} explores science and engineering cases beyond the baseline technology demonstration. These studies require simulations that are both realistic and easy to vary: one may wish to change the observing mode, probe sequence, estimator, controller settings, or image-generation fidelity while preserving enough common structure to make the results comparable, and eventually applicable on the real instrument. In particular, a higher-fidelity image is needed to study how the reference HOWFSC workflow behaves when the measurement images are more flight-like than the compact model used internally.

The \texttt{corgihowfsc} package addresses this need by placing a configuration-driven experiment layer around the Roman CGI tools. It keeps \texttt{cgi-howfsc} as the reference estimation and control tool, including compact-model Jacobian generation, while allowing probe images to be generated either by the compact model or by the higher-fidelity \texttt{corgisim} model \cite{Krist2023, Zhang2026}. The goal is not to replace the underlying optical or control packages, but to make controlled HOWFSC experiments easier to define, execute, inspect, and repeat. This enables repeatable studies of instrument performance, model mismatch, alternative probes, estimator choices, control settings, and runtime scaling.

\section{\texttt{corgihowfsc} Software Framework}
The \texttt{corgihowfsc} project is a CPP-developed wrapper and configurable experiment layer for Roman CGI HOWFSC simulations. It adds the machinery needed to run repeatable studies with selectable image generation, exposure planning, diagnostics, organised outputs, and local or multi-node parallel execution. 

The motivation for this framework is the need to connect two modelling regimes. The first is the compact model from \texttt{cgi-howfsc}, which is a simplified Fourier optics model that propagates through the coronagraph focal and pupil planes. Wavefront aberrations are derived through phase retrieval on the full model, then projected onto relevant planes. Because the compact model is computationally efficient, it is always used by the HOWFSC/GITL workflow in \texttt{cgi-howfsc} for Jacobian calculation and is the default method to simulate images. The second is the full model from \texttt{corgisim}, which propagates through the full Roman CGI optical train, including expected wavefront errors on individual optics, alignment degrees of freedom, and detector effects. The key design point is that \texttt{corgihowfsc} changes how images are simulated and how studies are configured, while preserving the reference sensing and control implementation used by the GITL loop.

\begin{figure}[!h]
    \centering
    \includegraphics[width=.9\textwidth]{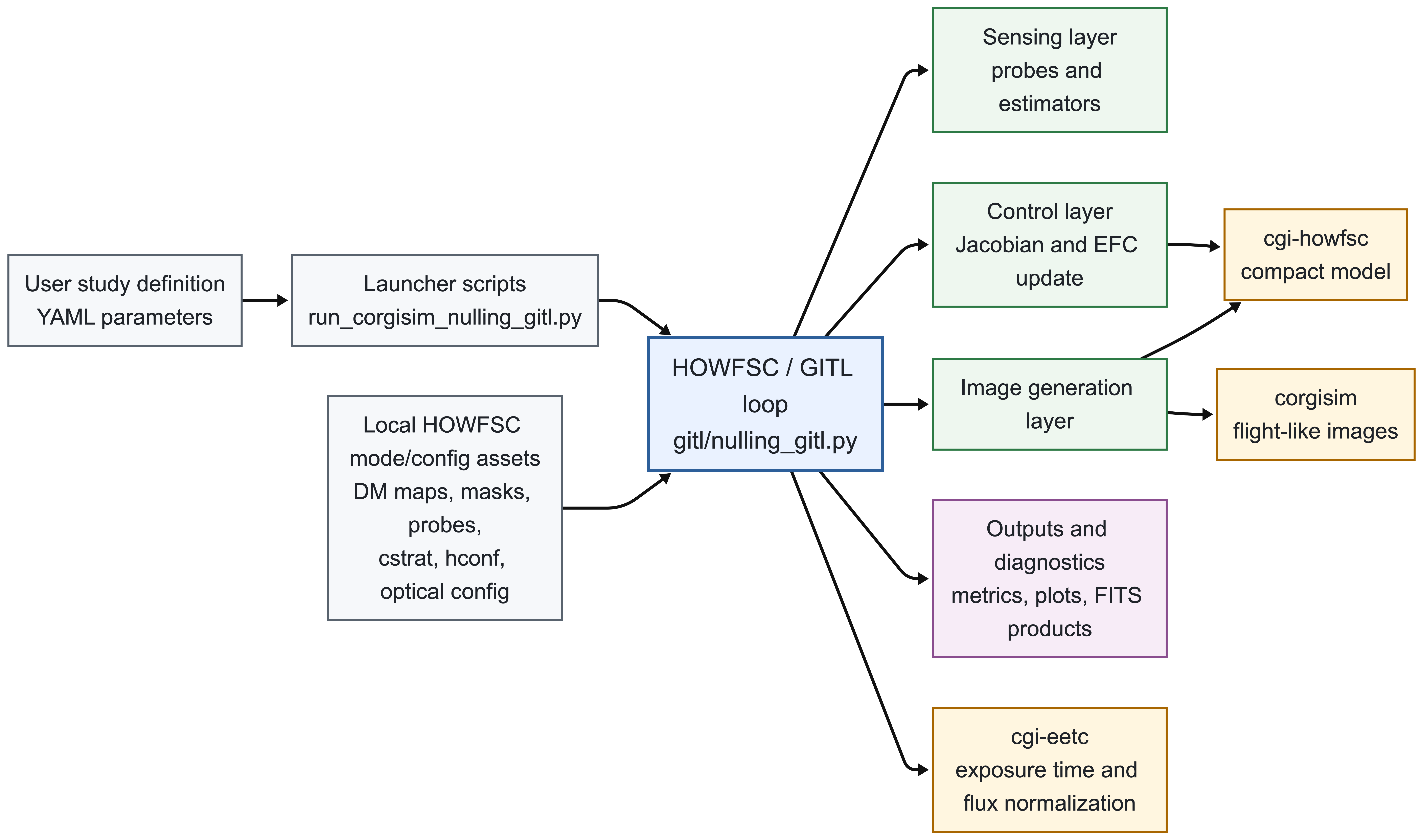 }\caption{Software architecture of \texttt{corgihowfsc}. The framework preserves the reference Roman CGI GITL/HOWFSC control path while exposing YAML-driven experiment setup, selectable image generation, compact-model Jacobian generation, exposure planning, camera settings, contrast normalisation, diagnostics, organised outputs, and local or parallel execution. 
}
    \label{fig:corgihowfsc-overview}
\end{figure}

\subsection{Functional architecture}
The framework can be viewed as four functional parts, summarised in Fig.~\ref{fig:corgihowfsc-overview}. The configuration part initialises the Roman CGI optical model and defines the settings for the selected observing mode, including the hardware configuration, DH geometry, probe sequence, estimator, controller, DM settings, exposure strategy, normalisation method, and outputs. The image-generation part selects how simulated images are produced: the compact model from \texttt{cgi-howfsc} for fast iteration, or higher-fidelity image generation through \texttt{corgisim} when more flight-like images are required. The reference GITL/HOWFSC part passes the resulting images through the sensing and control path, including the selected probes, estimator, compact-model Jacobian, and controller. The execution and diagnostics part manages local multiprocessing or multi-node distributed execution and writes a structured output directory containing the configuration, logs, images, metrics, DM commands, and diagnostic products needed to understand or reproduce the result.

This structure allows the same configured study to be run with different image-generation models or execution modes while keeping the underlying HOWFSC loop tied to the reference implementation. It is particularly useful when higher-fidelity image generation become computationally expensive, because the instrument configuration and the runtime strategy remain explicit parts of the same run record.

\subsection{Relationship to Roman CGI modelling tools}
This design keeps the official Roman CGI modelling packages in their intended roles. The \texttt{cgi-howfsc} code base provides the reference GITL/HOWFSC workflow, including the compact optical model used for Jacobian generation at all times. The \texttt{corgisim} code base provides higher-fidelity simulated images, including detector and the full optical train \cite{Krist2023} needed for more flight-like performance studies. The \texttt{cgi-eetc} \cite{cgieetc} code base provides the engineering exposure time calculator used to estimate peak fluxes, set exposure times, detector gain, and frame counts which allows us to convert images to contrast units, and track the expected observing time for each HOWFSC iteration. The \texttt{cgi-coralign} \cite{cgicoralign} code base provides calibration-related capabilities that can be incorporated into future or mode-specific workflows. Finally, the \texttt{corgihowfsc} code base ties these components together so that CPP studies can compare modelling assumptions and control choices without rewriting the underlying instrument models.

\section{GITL/HOWFSC Simulation Workflow}
A \texttt{corgihowfsc} run follows the GITL/HOWFSC pattern while allowing the measurement images to be generated either by the compact model or by \texttt{corgisim}. The HOWFSC experiment configuration is recorded in a YAML file, including the observing mode, image generation model, control settings, exposure and contrast conversion strategy, execution mode, and outputs. This makes the configuration file the primary record of the simulation while preserving the reference HOWFSC loop implementation.

\begin{verbatim}
active_model: corgihowfsc (corgisim)

sim_settings:
  mode: nfov_band1
  dark_hole: 360deg
  probe_shape: default
  precomp: precomp_jacs_always

models:
  corgihowfsc:
    backend_type: corgihowfsc
    normalization_type: eetc
    estimator: default
\end{verbatim}

In this example, \texttt{active\_model} selects the model block used for image generation, while \texttt{sim\_settings} and \texttt{models} define the observing mode, DH geometry, probe choice, Jacobian precomputation behaviour, normalisation strategy, and estimator. The HOWFSC estimator, controller, and compact-model Jacobian generation therefore remain tied to the \texttt{cgi-howfsc} workflow, even when the probe images supplied to the loop are generated by \texttt{corgisim}.

The \texttt{normalization\_type: eetc} setting selects the \texttt{cgi-eetc}-based strategy. Within the loop, \texttt{cgi-eetc} is used to compute peak fluxes for contrast conversion and to support the exposure time, detector gain, frame count, and iteration time estimates used when planning the next iteration.

\begin{figure}[!h]
    \centering
    \includegraphics[width=\textwidth]{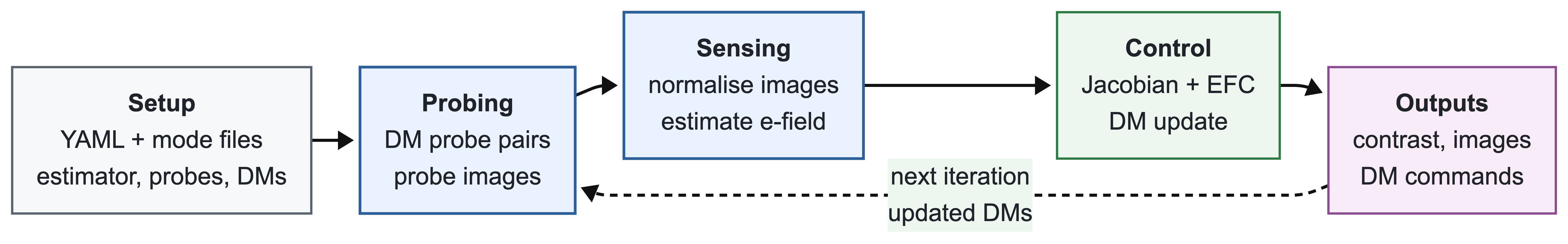}\caption{The \texttt{corgihowfsc} execution loop from configured image generation through contrast conversion, electric-field estimation, EFC control, image-acquisition planning, and diagnostic outputs.}
    \label{fig:corgihowfsc-loop}
\end{figure}

The execution loop, summarised in Fig.~\ref{fig:corgihowfsc-loop}, follows the standard HOWFSC pattern. For each control iteration, the selected image model produces the probed and unprobed science frames required by the estimator, in all selected wavelength bands. These frames are converted to contrast units, and pairwise probing is used to estimate the electric field over the controlled region. The controller then uses electric-field conjugation (EFC) \cite{Giveon2007SPIE}, together with the compact-model Jacobian, to compute the next DM command. This deliberate split between image generation and control is central to the framework: it allows \texttt{corgihowfsc} to test how the reference GITL/HOWFSC loop behaves when the images are more flight-like than the model used internally by the controller. After each iteration, \texttt{corgihowfsc} updates the acquisition plan and records the images, electric-field estimates, contrast metrics, DM commands, timing information, and other diagnostics requested by the run configuration.

This configuration-driven structure supports controlled comparisons between HOWFSC studies. The resulting output directories contain the metadata and diagnostics needed to trace which choices changes between runs and how those choices affected the simulated HOWFSC performance. 

\subsection{Image model selection}
The image model selection determines the fidelity and cost of image generation. Compact-model image generation is useful for rapid development, controller tuning, and comparisons to the baseline \texttt{cgi-howfsc} behaviour. Higher-fidelity \texttt{corgisim} image generation is used when the study requires a more realistic representation of the instrument response. In both cases, the present workflow uses the compact model for calculation of the control Jacobian.

\subsection{Probing, sensing and control loop}
The probing, sensing, and control settings define how the complex electric field is estimated and how the DM command are computed to reduce the coherent starlight in the DH. Probe choices determine the image diversity available to the estimator, while the controller uses the estimated field, compact-model Jacobian and regularisation choices to compute the next DM command. Keeping these settings explicit in the configuration makes parameter sweeps and method comparisons straightforward, including studies of new probe shapes, estimators, and control settings. The control loop behaviour is setup using a contrast strategy which enables, in the following examples, a beta-bumping strategy to enable convergence to deeper contrasts beyond local minima.

\subsection{Output organisation and diagnostics}
Each run is written to a self-contained output location. In addition to the images, the output products can include the input configuration, runtime logs, contrast histories, DM commands, estimator products, controller diagnostics, and summary figures. This organisation is intended to make each run inspectable by a human and reproducible by another user or computational environment.

\section{Example Study and Run Products}
\subsection{Run configuration}
For the present example, we use the Hybrid Lyot Coronagraph (HLC) narrow-field-of-view mode in Band~1 with a 360-degree dark hole. Images are generated with \texttt{corgisim}; the HOWFSC loop uses pairwise probing, EFC, and the configured estimator and controller settings from the YAML run file. The control Jacobian is still generated with the compact model, matching the reference \texttt{cgi-howfsc} control path. The same configuration can be executed locally for development or with multi-nodes for larger run with more iterations.

\begin{table}[!h]
\caption{Example configuration for a representative \texttt{corgihowfsc} closed-loop simulation run.}
\label{tab:example-config}
\begin{center}
\begin{tabular}{ll}
\hline
Setting & Example choice \\
\hline
Observing mode & Hybrid Lyot Coronagraph NFOV, Band~1 \\
Dark-hole geometry & 360-degree dark hole \\
Image model & \texttt{corgisim} \\
Control Jacobian & Compact model through \texttt{cgi-howfsc} \\
Wavefront sensing & Pairwise probing \\
Control method & Electric-field conjugation \\
Runtime & Local execution or MPI execution \\
\hline
\end{tabular}
\end{center}
\end{table}

\subsection{Output products}
\begin{figure}[!h]
    \centering
    \includegraphics[width=0.8\linewidth]{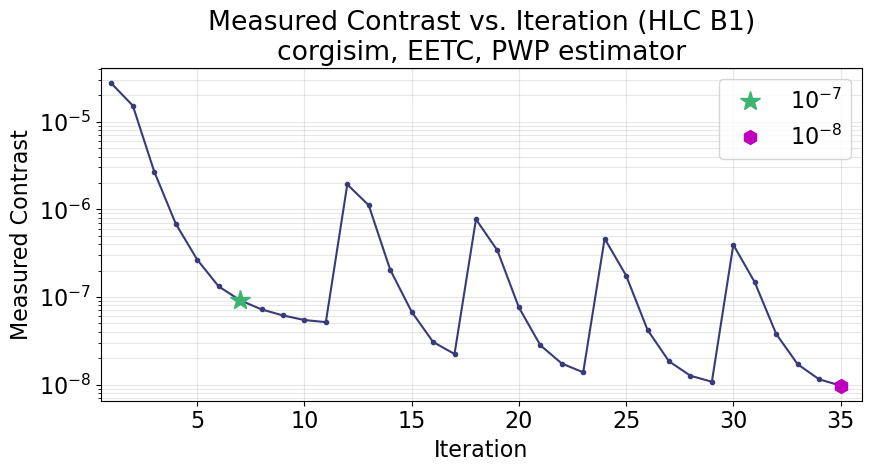}
    \caption{Measured mean DH contrast as a function of iteration for the representative \texttt{corgihowfsc} run.}
    \label{fig:contrast}
\end{figure}

\begin{figure}[!h]
    \centering
    \includegraphics[width=0.9\linewidth]{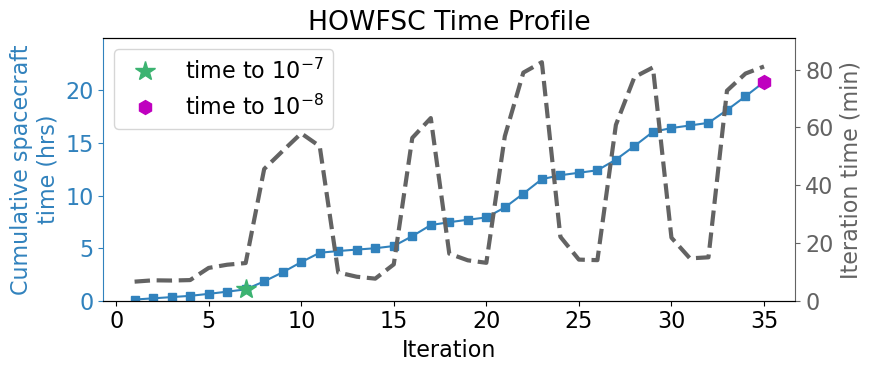}
    \caption{Cumulative time estimate for the representative HOWFSC run. The estimate includes the exposure time and number of frames requested for probed and unprobed images, but does not include GITL delays.}
    \label{fig:time_profile}
\end{figure}

For the example run, the primary scalar diagnostic is the measured dark-hole contrast as a function of HOWFSC iteration. Fig.~\ref{fig:contrast} tracks the iterations required to reach the target contrast levels of $10^{-7}$ and $10^{-8}$, and pairs this contrast history with a cumulative observing-time estimate that includes probe and unprobed image acquisitions in Fig.~\ref{fig:time_profile}. These two figures together separate HOWFSC convergence from observing efficiency: one shows how the contrast evolves, while the other shows the time cost of reaching the same contrast targets. The contrast versus iteration data is used to determine performance of a given set of HOWFSC parameters. We look at the number of iterations required to reach a given contrast as well as the best contrast achieved.

One of the outputs from \texttt{corgihowfsc} contains the time required by the spacecraft to complete an iteration based on the exposure time and number of frames requested for the probed and unprobed images. This allows us to determine the time required to reach a particular contrast. We note that Fig.~\ref{fig:time_profile}, the delays from the GITL architecture are not included.

\begin{figure}[!h]
    \centering
    \includegraphics[width=.7\linewidth]{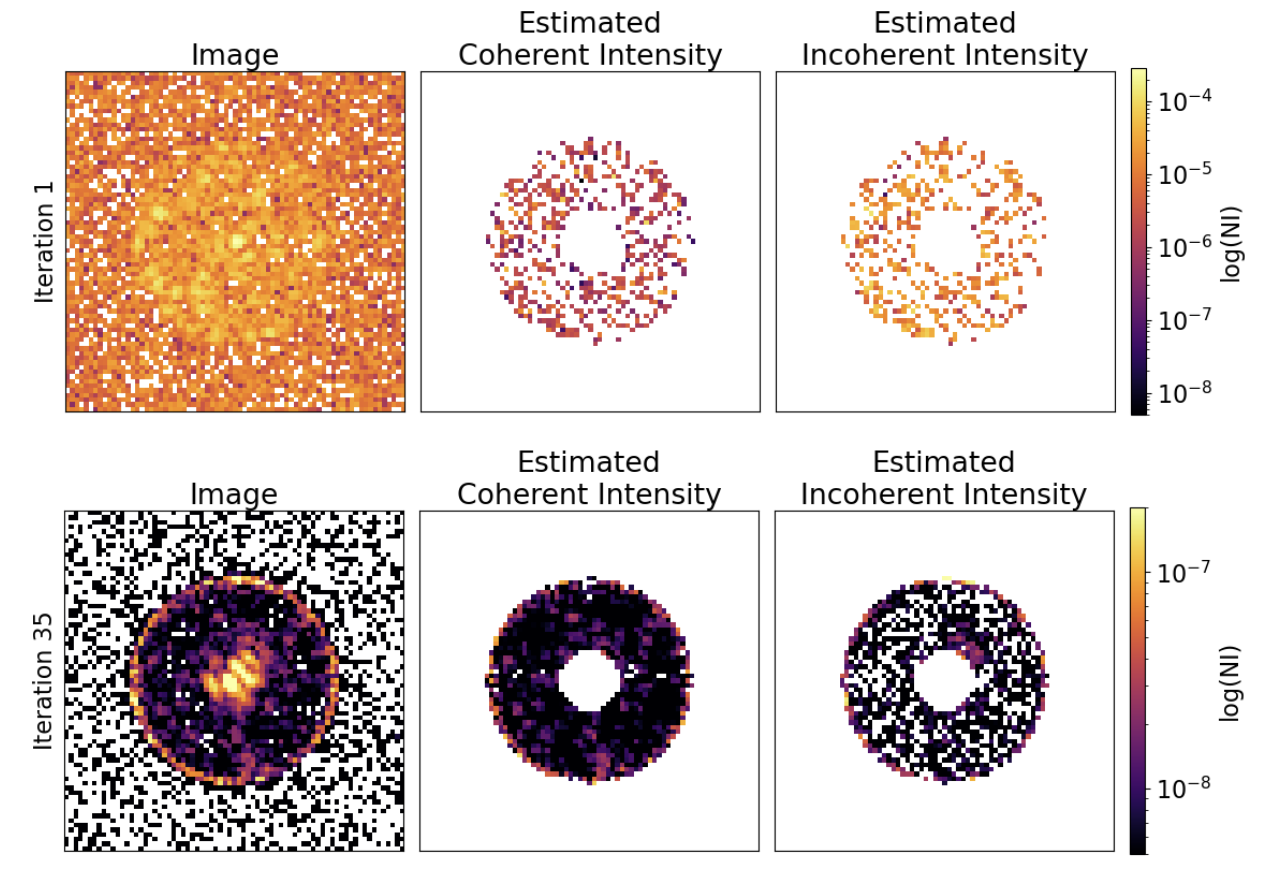}
    \caption{Example diagnostic image from the HOWFSC run in HLC Band 1, with pairwise probing (PWP) for electric field estimation and \texttt{corgisim} for image generation. The columns show the simulated image, estimated coherent intensity, and the estimated incoherent intensity at iteration 1 and iteration 35.}
    \label{fig:sample_frames}
\end{figure}

A second diagnostic checks the estimator by comparing coherent and incoherent light in Fig.~\ref{fig:sample_frames}. Pairwise probing estimates the coherent electric field, $E$, and the coherent intensity, $|E|^2$. The difference between the unprobed image and the estimated coherent intensity is labelled as the estimated incoherent intensity. In this context, ``incoherent'' describes light that is not captured by the estimator, potentially because it is not measurably modulated by the deformable mirrors.

\subsection{Diagnostic and saved products}
The diagnostic products are part of the scientific value of the package. They allow the user to check whether a run failed because of a modelling choice, a control parameter, an estimator behaviour, or an unexpected optical state. Useful products for this section include example science frames, probe images, estimated electric-field maps, contrast-history plots, DM-command maps, and a example directory tree showing the saved outputs. A typical output directory contains the run configuration, GITL log, contrast histories, generated FITS products, and per-iteration frames. The exact products depend on the requested diagnostics, here we use the following organisation:

% \begin{verbatim}
% run_directory/
% |-- config.yml
% |-- gitl.log
% |-- contrast_vs_iteration.pdf
% |-- measured_contrast.csv
% |-- predicted_contrast.csv
% |-- debugging_history.csv
% |-- final_frames.fits
% |-- iteration_XXXX/
%     |-- *.fits
% \end{verbatim}

\vspace{-1\baselineskip}
\begin{Verbatim}[fontsize=\footnotesize,
    baselinestretch=0.82]
run_directory/
|-- config.yml
|-- gitl.log
|-- contrast_vs_iteration.pdf
|-- measured_contrast.csv
|-- predicted_contrast.csv
|-- debugging_history.csv
|-- final_frames.fits
|-- iteration_XXXX/*.fits
\end{Verbatim}
\vspace{-1\baselineskip}

\section{Runtime and Scalability}
The computational cost of a HOWFSC study strongly depends on the selected image model. Compact model studies can often be performed on a workstation and are appropriate for rapid iteration. The \texttt{corgisim} image generation increases the cost of each frame and can make parameter sweeps expensive. The \texttt{corgihowfsc} package therefore includes runtime options that let the same configured experiment run locally using multiprocessing on a single node, or use the Message Passing Interface (MPI) to distribute independent image generation and Jacobian computation tasks across multiple nodes. 

\begin{figure}[!h]
    \centering
    \includegraphics[width=0.5\linewidth]{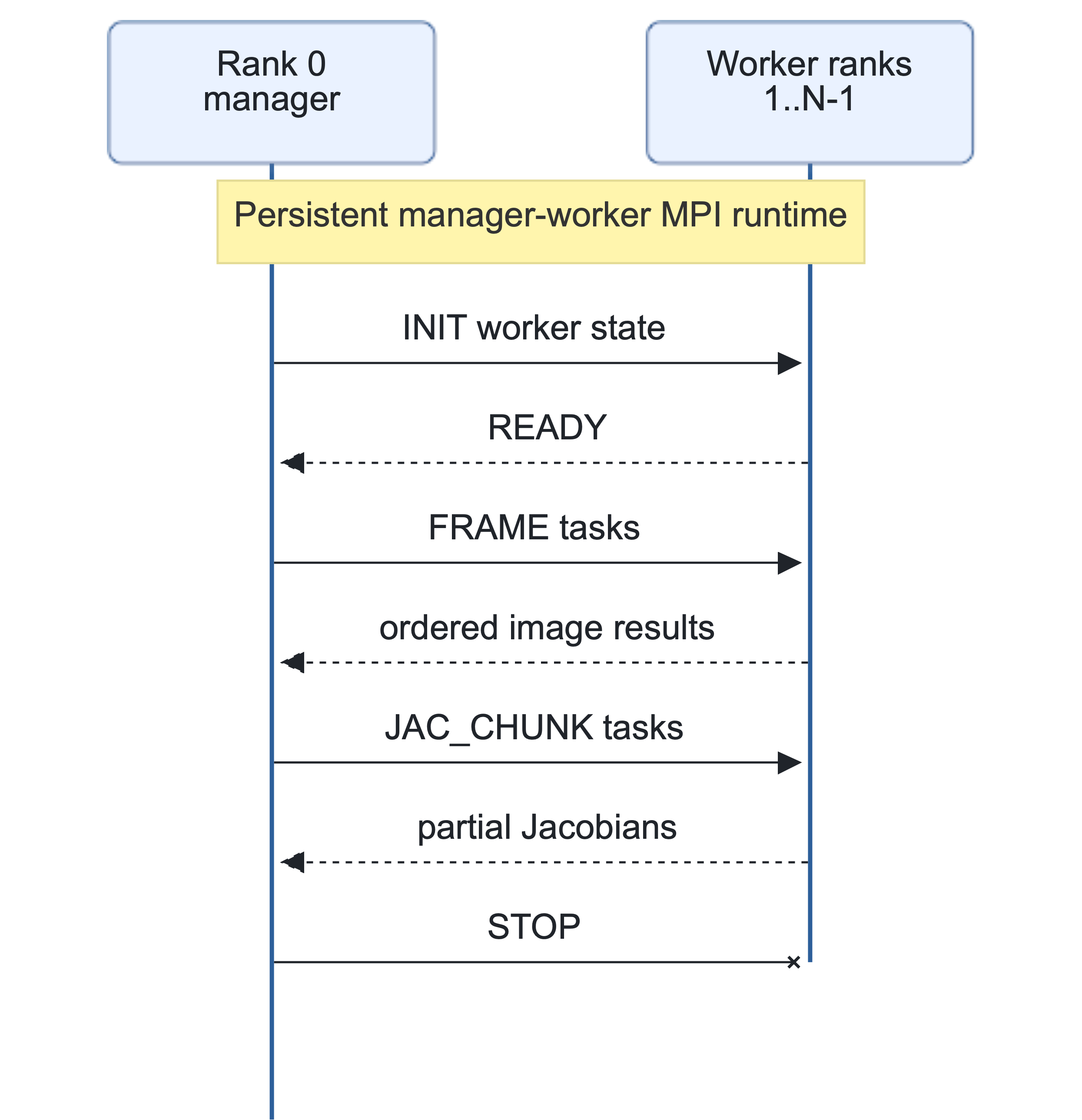}
    \caption{MPI manager--worker communication pattern. Rank 0 initialises persistent worker ranks, sends frame and Jacobian-chunk tasks, and receives ordered results for use in the HOWFSC loop. The manager terminates the workers after the HOWFSC loop.}
    \label{fig:mpi}
\end{figure}

MPI is a standard framework for coordinating work across multiple processes or compute nodes. In \texttt{corgihowfsc}, the MPI execution mode follows a manager-worker pattern: the manager process initialises the experiment, distributes independent frame generation or Jacobian-computation chunks to worker ranks, collects the completed products in a deterministic order, and passes them back to the HOWFSC loop, as shown in Fig.~\ref{fig:mpi}. The workers remain active throughout the HOWFSC loop and terminate only after receiving the final \texttt{STOP} message. Implementation details and message formats are provided in the \texttt{corgihowfsc} documentation. MPI keeps the HOWFSC configuration unchanged while changing only the execution backend, which is useful when moving from a laptop-scale test to a cluster-scale parameter sweep.

\begin{figure}[!h]
    \centering
    \includegraphics[width=\linewidth]{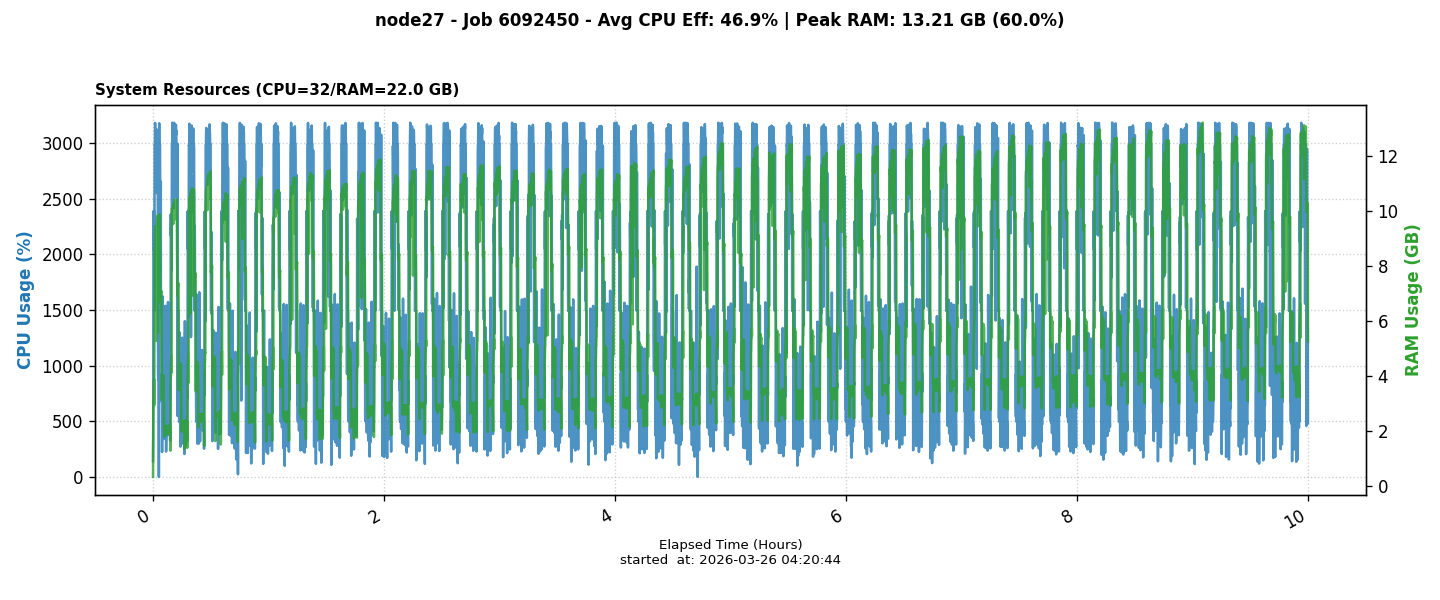}
    \caption{Resources usage during a 70-iteration HOWFSC run using \texttt{corgisim} image generation without MPI. The blue curve shows the combine CPU usage across 32 allocated CPU cores (left axis), where 3200 \% corresponds to full utilisation of all cores. The green curves shows the RAM usage in GB (right axis). The repeated peaks correspond to the image gerneration stages of each HOWFSC iteration, separated by periods lower utilisation stages. The run completed in approxiamtely 10 hours, with an average CPU efficiency of 46.9 \% and a peak memory of 13.21~GB, corresponding to 60\% of the allocated 22~GB.}
    \label{fig:runtime_parallel_process_wo_mpi}
\end{figure}

Figure \ref{fig:runtime_parallel_process_wo_mpi} shows the resource usage and elapsed time of a representative HOWFSC run using \texttt{corgisim} image generation without MPI. The total wall-clock time is dominated by the image generation, which scales with the number of frames requested and number of process needed for image generation. The MPI manager--worker pattern is designed to reduce the wall-clock time by distributing the image generation and Jacobian computation across multiple nodes. 

\section{Current Limitations and Future Work}
The current framework is designed to make realistic CPP HOWFSC studies repeatable. Near-term work includes systematic comparisons between compact-model and \texttt{corgisim}-generated images, parameter sweeps over probe designs and controller regularisation, and representative HOWFSC use cases that exercise different observing modes. Additional work should clarify which products are required for each mode and how those products should be stored with the simulation outputs. The package should continue to be checked against baseline \texttt{cgi-howfsc} behaviour for compact-model runs, then extended to document where higher-fidelity image generation changes the estimated field, contrast per iteration, or DM commands. Runtime benchmarks on local workstations and MPI clusters will also help users choose an appropriate execution mode for each study.

\section{Summary}
The \texttt{corgihowfsc} package provides a configuration-driven framework for Roman CGI HOWFSC simulations. It connects the baseline \texttt{cgi-howfsc} control workflow with higher-fidelity image generation, exposure planning, camera settings, contrast normalisation, calibration-related tools, structured diagnostics, and local or distributed execution. This structure allows CPP teams to run repeatable studies in which observing modes, image models, probes, estimators, controllers, and DM settings can be varied independently. The resulting workflow supports both rapid compact-model development and higher-fidelity simulations needed to evaluate future Roman CGI science and engineering investigations. Further details on the implementation, configuration options and usage are provided in the package documentation \footnote{\url{https://roman-corgi.github.io/corgihowfsc/}}. 

\acknowledgments 
The research was carried out in part through the Jet Propulsion Laboratory, California Institute of Technology, under a contract with the National Aeronautics and Space Administration (80NM0018D0004). A. Lau acknowledges the support by the European Union (ERC, ESCAPE, project No. 101044152). Views and opinions expressed are, however, those of the author(s) only and do not necessarily reflect those of the European Union or the European Research Council Executive Agency. This research has made use of computing facilities operated by the CeSAM data centre at LAM, Marseille, France. Large language models were utilised for code debugging, plot optimisation and language editing in this work. This work made use of \texttt{numpy}~\cite{harris2020array}, \texttt{scipy}~\cite{virtanen2020scipy}, \texttt{matplotlib}~\cite{hunter2007matplotlib}, \texttt{astropy}~\cite{astropy2022} and \texttt{pytest}~\cite{pytest}.

% References
\nocite{*}
\bibliography{report} % bibliography data in report.bib

@INPROCEEDINGS{Bailey2023,
       author = {{Bailey}, Vanessa P. and {Bendek}, Eduardo and {Monacelli}, Brian and {Baker}, Caleb and {Bedrosian}, Gasia and {Cady}, Eric and {Douglas}, Ewan S. and {Groff}, Tyler and {Hildebrandt}, Sergi R. and {Kasdin}, N. Jeremy and {Krist}, John and {Macintosh}, Bruce and {Mennesson}, Bertrand and {Morrissey}, Patrick and {Poberezhskiy}, Ilya and {Subedi}, Hari B. and {Rhodes}, Jason and {Roberge}, Aki and {Ygouf}, Marie and {Zellem}, Robert T. and {Zhao}, Feng and {Zimmerman}, Neil T.},
        title = "{Nancy Grace Roman Space Telescope coronagraph instrument overview and status}",
    booktitle = {Society of Photo-Optical Instrumentation Engineers (SPIE) Conference Series},
         year = 2023,
       series = {Society of Photo-Optical Instrumentation Engineers (SPIE) Conference Series},
       volume = {12680},
        month = oct,
          eid = {126800T},
        pages = {126800T},
          doi = {10.1117/12.2679036},
archivePrefix = {arXiv},
       eprint = {2309.08672},
 primaryClass = {astro-ph.IM},
       adsurl = {https://ui.adsabs.harvard.edu/abs/2023SPIE12680E..0TB}
}

@INPROCEEDINGS{Savransky2024,
       author = {{Savransky}, Dmitry and {Bailey}, Vanessa P. and {Wolff}, Schuyler G. and {Millar-Blanchaer}, Maxwell A. and {Wang}, Jason and {Altinier}, Lisa and {Anche}, Ramya and {Baudoz}, Pierre and {Biller}, Beth and {Blunt}, Sarah and {Brandner}, Wolfgang and {Brinjikji}, Marah and {Carri{\'o}n-Gonz{\'a}lez}, Oscar and {Chavez}, Amanda and {Choquet}, Elodie and {Doelman}, David and {Girard}, Julien H. and {Greenbaum}, Alexandra Z. and {Hasler}, Samantha N. and {Hom}, Justin and {Ingalls}, James G. and {Kane}, Stephen R. and {Kasdin}, N. Jeremy and {Krause}, Oliver and {Kuzuhara}, Masayuki and {Lau}, Alexis and {Li}, Zhexing and {Livingston}, John and {Lowrance}, Patrick J. and {Ludwick}, Kevin and {Macintosh}, Bruce and {Mamajek}, Eric and {Marley}, Mark and {Mazoyer}, Johan and {Mennesson}, Bertrand and {Mizuki}, Toshiyuki and {Moran}, Sarah E. and {Murakami}, Naoshi and {Nishikawa}, Jun and {Noel}, Malachi and {Pueyo}, Laurent and {Hildebrandt Rafels}, Sergi and {Rhodes}, Jason and {Robinson}, Tyler and {De Rosa}, Robert J. and {Samland}, Matthias and {Schragal}, Nicholas and {Schreiber}, J{\"u}rgen and {Sobeck}, Jennifer and {Stapelfeldt}, Karl and {Tamura}, Motohide and {Uyajma}, Taichi and {Vigan}, Arthur and {Woodland}, Michele and {Ygouf}, Marie and {Yoneta}, Kenta and {Zellem}, Robert T. and {Zimmerman}, Neil T.},
        title = "{The Nancy Grace Roman Space Telescope coronagraph community participation program}",
    booktitle = {Space Telescopes and Instrumentation 2024: Optical, Infrared, and Millimeter Wave},
         year = 2024,
       editor = {{Coyle}, Laura E. and {Matsuura}, Shuji and {Perrin}, Marshall D.},
       series = {Society of Photo-Optical Instrumentation Engineers (SPIE) Conference Series},
       volume = {13092},
        month = aug,
          eid = {130921I},
        pages = {130921I},
          doi = {10.1117/12.3020514},
       adsurl = {https://ui.adsabs.harvard.edu/abs/2024SPIE13092E..1IS}
}

@INPROCEEDINGS{Millar-Blanchaer2024,
       author = {{Millar-Blanchaer}, Maxwell Andrew and {Wang}, Jason and {Bogat}, Ellis and {Schreiber}, J{\"u}rgen and {Ygouf}, Marie and {Ludwick}, Kevin J. and {Greenbaum}, Alexandra Z. and {Bailey}, Vanessa and {Hildebrandt Rafels}, Sergi and {Savransky}, Dmitry and {Samland}, Matthias and {Altinier}, Lisa and {Anche}, Ramya and {Biller}, Beth and {Chavez}, Amanda and {Choquet}, Elodie and {Girard}, Julien H. and {Hom}, Justin and {Ingalls}, James G. and {Kasdin}, N. Jeremy and {Krause}, Oliver and {Livingston}, John and {Mazoyer}, Johan and {Pueyo}, Laurent and {Uyama}, Taichi and {Zellem}, Robert T. and {Zimmerman}, Neil T.},
        title = "{The Roman coronagraph community participation program: data reduction and simulations}",
    booktitle = {Space Telescopes and Instrumentation 2024: Optical, Infrared, and Millimeter Wave},
         year = 2024,
       editor = {{Coyle}, Laura E. and {Matsuura}, Shuji and {Perrin}, Marshall D.},
       series = {Society of Photo-Optical Instrumentation Engineers (SPIE) Conference Series},
       volume = {13092},
        month = aug,
          eid = {1309256},
        pages = {1309256},
          doi = {10.1117/12.3020478},
       adsurl = {https://ui.adsabs.harvard.edu/abs/2024SPIE13092E..56M}
}

@conference{Zhang2026,
  title={The Roman Coronagraph Community Participation Program: “corgisim” — a simulation suite for the Nancy Grace Roman Space Telescope Coronagraphic Instrument},
 author={Jingwen Zhang and Sophie Noiret and Maxwell A. Millar-Blanchaer and Jason J. Wang and Neil T. Zimmerman and Eric Shen and Alexis Lau and Elodie Choquet and Taichi Uyajma and Chen Xie and Jessica Gersh-Range and Julien H. Girard and Vanessa P. Bailey and John E. Krist and Kevin J. Ludwick and Alexis Bidot and Arthur
Vigan},
    booktitle={This Conference},
    year={2026},
}

@ARTICLE{astropy2022,
               author = {{Astropy Collaboration} and {Price-Whelan}, Adrian M. and {Lim}, Pey Lian and {Earl}, Nicholas and {Starkman}, Nathaniel and {Bradley}, Larry and {Shupe}, David L. and {Patil}, Aarya A. and {Corrales}, Lia and {Brasseur}, C.~E. and {N{"o}the}, Maximilian and {Donath}, Axel and {Tollerud}, Erik and {Morris}, Brett M. and {Ginsburg}, Adam and {Vaher}, Eero and {Weaver}, Benjamin A. and {Tocknell}, James and {Jamieson}, William and {van Kerkwijk}, Marten H. and {Robitaille}, Thomas P. and {Merry}, Bruce and {Bachetti}, Matteo and {G{"u}nther}, H. Moritz and {Aldcroft}, Thomas L. and {Alvarado-Montes}, Jaime A. and {Archibald}, Anne M. and {B{'o}di}, Attila and {Bapat}, Shreyas and {Barentsen}, Geert and {Baz{'a}n}, Juanjo and {Biswas}, Manish and {Boquien}, M{'e}d{'e}ric and {Burke}, D.~J. and {Cara}, Daria and {Cara}, Mihai and {Conroy}, Kyle E. and {Conseil}, Simon and {Craig}, Matthew W. and {Cross}, Robert M. and {Cruz}, Kelle L. and {D'Eugenio}, Francesco and {Dencheva}, Nadia and {Devillepoix}, Hadrien A.~R. and {Dietrich}, J{"o}rg P. and {Eigenbrot}, Arthur Davis and {Erben}, Thomas and {Ferreira}, Leonardo and {Foreman-Mackey}, Daniel and {Fox}, Ryan and {Freij}, Nabil and {Garg}, Suyog and {Geda}, Robel and {Glattly}, Lauren and {Gondhalekar}, Yash and {Gordon}, Karl D. and {Grant}, David and {Greenfield}, Perry and {Groener}, Austen M. and {Guest}, Steve and {Gurovich}, Sebastian and {Handberg}, Rasmus and {Hart}, Akeem and {Hatfield-Dodds}, Zac and {Homeier}, Derek and {Hosseinzadeh}, Griffin and {Jenness}, Tim and {Jones}, Craig K. and {Joseph}, Prajwel and {Kalmbach}, J. Bryce and {Karamehmetoglu}, Emir and {Ka{l}uszy{'n}ski}, Miko{l}aj and {Kelley}, Michael S.~P. and {Kern}, Nicholas and {Kerzendorf}, Wolfgang E. and {Koch}, Eric W. and {Kulumani}, Shankar and {Lee}, Antony and {Ly}, Chun and {Ma}, Zhiyuan and {MacBride}, Conor and {Maljaars}, Jakob M. and {Muna}, Demitri and {Murphy}, N.~A. and {Norman}, Henrik and {O'Steen}, Richard and {Oman}, Kyle A. and {Pacifici}, Camilla and {Pascual}, Sergio and {Pascual-Granado}, J. and {Patil}, Rohit R. and {Perren}, Gabriel I. and {Pickering}, Timothy E. and {Rastogi}, Tanuj and {Roulston}, Benjamin R. and {Ryan}, Daniel F. and {Rykoff}, Eli S. and {Sabater}, Jose and {Sakurikar}, Parikshit and {Salgado}, Jes{'u}s and {Sanghi}, Aniket and {Saunders}, Nicholas and {Savchenko}, Volodymyr and {Schwardt}, Ludwig and {Seifert-Eckert}, Michael and {Shih}, Albert Y. and {Jain}, Anany Shrey and {Shukla}, Gyanendra and {Sick}, Jonathan and {Simpson}, Chris and {Singanamalla}, Sudheesh and {Singer}, Leo P. and {Singhal}, Jaladh and {Sinha}, Manodeep and {Sip{H{o}}cz}, Brigitta M. and {Spitler}, Lee R. and {Stansby}, David and {Streicher}, Ole and {{{S}}umak}, Jani and {Swinbank}, John D. and {Taranu}, Dan S. and {Tewary}, Nikita and {Tremblay}, Grant R. and {Val-Borro}, Miguel de and {Van Kooten}, Samuel J. and {Vasovi{'c}}, Zlatan and {Verma}, Shresth and {de Miranda Cardoso}, Jos{'e} Vin{'i}cius and {Williams}, Peter K.~G. and {Wilson}, Tom J. and {Winkel}, Benjamin and {Wood-Vasey}, W.~M. and {Xue}, Rui and {Yoachim}, Peter and {Zhang}, Chen and {Zonca}, Andrea and {Astropy Project Contributors}},
                title = "{The Astropy Project: Sustaining and Growing a Community-oriented Open-source Project and the Latest Major Release (v5.0) of the Core Package}",
              journal = {Astrophysical Journal},
                 year = 2022,
                month = aug,
               volume = {935},
               number = {2},
                  eid = {167},
                pages = {167},
                  doi = {10.3847/1538-4357/ac7c74},
        archivePrefix = {arXiv},
               eprint = {2206.14220},
         primaryClass = {astro-ph.IM},
               adsurl = {https://ui.adsabs.harvard.edu/abs/2022ApJ...935..167A}
        }

@misc{pytest,
  author       = {Holger Krekel and Bruno Oliveira and Ronny Pfannschmidt and Floris Bruynooghe and Brianna Laugher and Freya Bruhin},
  title        = {pytest x.y},
  year         = {2004},
  howpublished = {\url{https://github.com/pytest-dev/pytest}},
  note         = {Version x.y. Contributors include Holger Krekel, Bruno Oliveira, Ronny Pfannschmidt, Floris Bruynooghe, Brianna Laugher, Freya Bruhin, and others.}
}

@software{corgidrp,
  author       = {Jason Wang and
                  juliamilton and
                  JuergenSchreiber and
                  Max Millar-Blanchaer and
                  manduhmia and
                  everlastingEric and
                  Aneesh Baburaj and
                  Ramya Anche and
                  Toshiyuki Mizuki and
                  Taichi Uyama and
                  Neil Zimmerman and
                  Alexis Lau and
                  Kevin Ludwick and
                  William O Balmer and
                  Matthias Samland and
                  Ben Sutlieff and
                  adrienmaillard and
                  Clarissa Rizzo Credidio Do O and
                  A J Eldorado Riggs and
                  JingwenZ and
                  Michael Liang and
                  Jessica Gersh-Range and
                  Lisa Altinier and
                  Giovanni Maria Strampelli and
                  John Livingston and
                  Kylee Renee Fluckiger and
                  mygouf and
                  Dmitry Savransky and
                  Ell Bogat and
                  TsutsumiNagai},
  title        = {roman-corgi/corgidrp: v5.0},
  month        = jun,
  year         = 2026,
  publisher    = {Zenodo},
  version      = {v5.0},
  doi          = {10.5281/zenodo.20650643},
  url          = {https://doi.org/10.5281/zenodo.20650643},
}

@ARTICLE{Cady2025,
       author = {{Cady}, Eric and {Bowman}, Nicholas and {Greenbaum}, Alexandra Z. and {Ingalls}, James G. and {Kern}, Brian and {Krist}, John and {Marx}, David and {Poberezhskiy}, Ilya and {Eldorado Riggs}, A.~J. and {Ruane}, Garreth and {Seo}, Byoung-Joon and {Shi}, Fang and {Zhou}, Hanying},
        title = "{High-order wavefront sensing and control for the Roman Coronagraph Instrument (CGI): architecture and measured performance}",
      journal = {Journal of Astronomical Telescopes, Instruments, and Systems},
         year = 2025,
        month = apr,
       volume = {11},
          eid = {021408},
        pages = {021408},
          doi = {10.1117/1.JATIS.11.2.021408},
archivePrefix = {arXiv},
       eprint = {2507.23738},
 primaryClass = {astro-ph.IM},
       adsurl = {https://ui.adsabs.harvard.edu/abs/2025JATIS..11b1408C}
}

@INPROCEEDINGS{Mennesson2022, 
       author = {{Mennesson}, B. and {Bailey}, V.~P. and {Zellem}, R. and {Hildebrandt}, S. and {Ygouf}, M. and {Rhodes}, J. and {Zimmerman}, N. and {Nemati}, B. and {Gonzalez}, G. and {Cady}, E. and {Kern}, B. and {Koch}, T. and {Krist}, J. and {Heydorff}, K. and {Luchik}, T. and {Mok}, F. and {Morrissey}, P. and {Poberezhskiy}, I. and {Riggs}, A.~J. and {Shi}, F. and {Zhao}, F. and {Akeson}, R. and {Armus}, L. and {Greenbaum}, A. and {Ingalls}, J. and {Lowrance}, P.},
        title = "{The Roman Space Telescope coronagraph technology demonstration: current status and relevance to future missions}",
    booktitle = {Space Telescopes and Instrumentation 2022: Optical, Infrared, and Millimeter Wave},
         year = 2022,
       editor = {{Coyle}, Laura E. and {Matsuura}, Shuji and {Perrin}, Marshall D.},
       series = {Society of Photo-Optical Instrumentation Engineers (SPIE) Conference Series},
       volume = {12180},
        month = aug,
          eid = {121801W},
        pages = {121801W},
          doi = {10.1117/12.2629176},
       adsurl = {https://ui.adsabs.harvard.edu/abs/2022SPIE12180E..1WM}
}

@ARTICLE{Krist2023,
       author = {{Krist}, John E. and {Steeves}, John B. and {Dube}, Brandon D. and {Eldorado Riggs}, A.~J. and {Kern}, Brian D. and {Marx}, David S. and {Cady}, Eric J. and {Zhou}, Hanying and {Poberezhskiy}, Ilya Y. and {Baker}, Caleb W. and {McGuire}, James P. and {Nemati}, Bijan and {Kuan}, Gary M. and {Mennesson}, Bertrand and {Trauger}, John T. and {Saini}, Navtej S. and {Rafels}, Sergi Hildebrandt},
        title = "{End-to-end numerical modeling of the Roman Space Telescope coronagraph}",
      journal = {Journal of Astronomical Telescopes, Instruments, and Systems},
         year = 2023,
        month = oct,
       volume = {9},
          eid = {045002},
        pages = {045002},
          doi = {10.1117/1.JATIS.9.4.045002},
archivePrefix = {arXiv},
       eprint = {2309.16012},
 primaryClass = {astro-ph.IM},
       adsurl = {https://ui.adsabs.harvard.edu/abs/2023JATIS...9d5002K}
}

@INPROCEEDINGS{Giveon2007SPIE,
       author = {{Give'on}, Amir and {Kern}, Brian and {Shaklan}, Stuart and {Moody}, Dwight C. and {Pueyo}, Laurent},
        title = "{Broadband wavefront correction algorithm for high-contrast imaging systems}",
    booktitle = {Astronomical Adaptive Optics Systems and Applications III},
         year = 2007,
       editor = {{Tyson}, Robert K. and {Lloyd-Hart}, Michael},
       series = {Society of Photo-Optical Instrumentation Engineers (SPIE) Conference Series},
       volume = {6691},
        month = sep,
          eid = {66910A},
        pages = {66910A},
          doi = {10.1117/12.733122},
       adsurl = {https://ui.adsabs.harvard.edu/abs/2007SPIE.6691E..0AG}
}

@INPROCEEDINGS{Giveon2011SPIE,
       author = {{Give'on}, Amir and {Kern}, Brian D. and {Shaklan}, Stuart},
        title = "{Pair-wise, deformable mirror, image plane-based diversity electric field estimation for high contrast coronagraphy}",
    booktitle = {Techniques and Instrumentation for Detection of Exoplanets V},
         year = 2011,
       editor = {{Shaklan}, Stuart},
       series = {Society of Photo-Optical Instrumentation Engineers (SPIE) Conference Series},
       volume = {8151},
        month = oct,
          eid = {815110},
        pages = {815110},
          doi = {10.1117/12.895117},
       adsurl = {https://ui.adsabs.harvard.edu/abs/2011SPIE.8151E..10G}
}

@software{cgihowfsc,
  author  = {Cady, Eric and Riggs, A. J. Eldorado and Marx, David
             and Bottom, Michael and Ludwick, Kevin},
  title   = {{cgi-howfsc}: High-order wavefront sensing and control
             for coronagraphic instruments and testbeds},
  year    = {2025},
  url     = {https://github.com/nasa-jpl/cgi-howfsc},
  urldate = {2026-07-17}
}

@software{cgieetc,
  author  = {Halverson, Sam and Miller, Sam and Ludwick, Kevin and Cady, Eric},
  title   = {{cgi-eetc}: CGI Engineering Exposure Time Calculator},
  year    = {2025},
  url     = {https://github.com/nasa-jpl/cgi-eetc},
  urldate = {2026-07-17}
}

@article{harris2020array,
  author = {Harris, Charles R. and Millman, K. Jarrod and
            van der Walt, St{\'e}fan J. and Gommers, Ralf and
            Virtanen, Pauli and Cournapeau, David and Wieser, Eric and
            Taylor, Julian and Berg, Sebastian and Smith, Nathaniel J. and
            Kern, Robert and Picus, Matti and Hoyer, Stephan and
            van Kerkwijk, Marten H. and Brett, Matthew and Haldane, Allan and
            Fern{\'a}ndez del R{\'i}o, Jaime and Wiebe, Mark and
            Peterson, Pearu and G{\'e}rard-Marchant, Pierre and
            Sheppard, Kevin and Reddy, Tyler and Weckesser, Warren and
            Abbasi, Hameer and Gohlke, Christoph and Oliphant, Travis E.},
  title = {Array programming with {NumPy}},
  journal = {Nature},
  year = {2020},
  volume = {585},
  number = {7825},
  pages = {357--362},
  doi = {10.1038/s41586-020-2649-2}
}

@article{virtanen2020scipy,
  author = {Virtanen, Pauli and Gommers, Ralf and Oliphant, Travis E. and
            Haberland, Matt and Reddy, Tyler and Cournapeau, David and
            Burovski, Evgeni and Peterson, Pearu and Weckesser, Warren and
            Bright, Jonathan and van der Walt, St{\'e}fan J. and
            Brett, Matthew and Wilson, Joshua and Millman, K. Jarrod and
            Mayorov, Nikolay and Nelson, Andrew R. J. and Jones, Eric and
            Kern, Robert and Larson, Eric and Carey, C. J. and
            Polat, {\.I}lhan and Feng, Yu and Moore, Eric W. and
            VanderPlas, Jake and Laxalde, Denis and Perktold, Josef and
            Cimrman, Robert and Henriksen, Ian and Quintero, E. A. and
            Harris, Charles R. and Archibald, Anne M. and
            Ribeiro, Ant{\^o}nio H. and Pedregosa, Fabian and
            van Mulbregt, Paul and {SciPy 1.0 Contributors}},
  title = {{SciPy} 1.0: Fundamental Algorithms for Scientific Computing
           in Python},
  journal = {Nature Methods},
  year = {2020},
  volume = {17},
  number = {3},
  pages = {261--272},
  doi = {10.1038/s41592-019-0686-2}
}

@article{hunter2007matplotlib,
  author = {Hunter, John D.},
  title = {Matplotlib: A 2D Graphics Environment},
  journal = {Computing in Science \& Engineering},
  year = {2007},
  volume = {9},
  number = {3},
  pages = {90--95},
  doi = {10.1109/MCSE.2007.55}
}

@inproceedings{cgicoralign,
  author    = {Riggs, A. J. Eldorado and
               Bertagna, Michael A. and
               Ruane, Garreth J. and
               Cady, Eric J. and
               Marx, David S. and
               Halverson, Samuel P. and
               Miller, Samuel and
               Ludwick, Kevin J.},
  title     = {{Coralign}: A Software Package for Coronagraphic
               Alignment and Calibration},
  booktitle = {Techniques and Instrumentation for Detection of Exoplanets XI},
  editor    = {Ruane, Garreth J.},
  volume    = {12680},
  pages     = {126802F},
  publisher = {SPIE},
  year      = {2023},
  doi       = {10.1117/12.2677703},
  url       = {https://doi.org/10.1117/12.2677703}
}
\bibliographystyle{spiebib} % makes bibtex use spiebib.bst
\end{document}